22 April 2013

Editor-in-Chief,
*arXiv*

Dear Sir:

Enclosed please find the original Word files of a new paper entitled "The atomic-level mechanism underlying the functionality of aquaporin-0".

So far, more than 82,000 protein structures have been reported in the Protein Data Bank, but the driving force and structures that allow for protein functions have not been elucidated at the atomic level for even one protein. We have found that the inter-subunit hydrophobic interaction drove the electrostatic opening of the aquaporin pore. Furthermore, this paper contains a video clearly showing the opening, closing and re-opening of the pore. We have included a simple schematic of an AQPO atomic machine as figure 1, as the three-dimensional structure is complex.

We are confident that our work is ground-breaking, and suitable for publication in *arXiv, quantative biology, biomolecules*. The paper has been edited by native English speakers, but the verbal description is complex and viewing video in the supporting information is in fact more convincing than the detailed description of the chemistry involved.

Please address all communications to me at: Department of Molecular Pharmacology and Biological Chemistry, Northwestern University Medical School, Chicago, IL 60611-3008, USA. E-mail: k-goto@northwestern.edu

We greatly appreciate your action on this matter, and look forward to hearing from you at your earliest convenience.

Sincerely yours,

Kunihiko Goto, M.D., Ph.D.
Research Professor

# The atomic-level mechanism underlying the functionality of aquaporin-0

**Atsushi Suenaga**[1], **Takehiko Ogura**[2], **Makoto Taiji**[1], **Akira Toyama**[3], **Hideo Takeuchi**[4], **Mingyu Son**[5], **Kazuyoshi Takayama**[5], **Masatoshi Iwamoto**[6], **Ikuro Sato**[7], **Jay Z. Yeh**[8], **Toshio Narahashi**[8], **Haruaki Nakaya**[2], **Akihiko Konagaya**[1] & **Kunihiko Goto**[8]*

[1]Bioinformatics Group, RIKEN Genomic Sciences Center, 61-1 Ono, Tsurumi, Yokohama, Kanagawa 230-0046 Japan.
[2]Department of Pharmacology, Chiba University Graduate School of Medicine, Chiba, Japan.
[3]Division of Pharmacy, Medical and Dental Hospital, Niigata University, 1-754 Asahimachi-dori, Niigata 951-8520, Japan.
[4]Department of Pharmaceuticals, Graduate School of Pharmaceutical Sciences, Tohoku University, Aobayama, Sendai 980-8578, Japan.
[5]Shock Wave Research Center, Institute of Fluid Science, Tohoku University, 2-1-1, Katahira, Sendai 980-8577, Japan.
[6]Department of Applied Physics, Tohoku Gakuin University, 1-13-1, Tagajyo, Miyagi 985-8537, Japan.
[7]Miyagi Prefecture Cancer Center, Natori, Miyagi, Japan.
[8]Department of Molecular Pharmacology and Biological Chemistry, Northwestern University Medical School, 303 East Chicago Avenue, Chicago, IL 60611-3008, USA.

*To whom correspondence should be addressed. E-mail: k-goto@northwestern.edu

## Abstract

So far, more than 82,000 protein structures have been reported in the Protein Data Bank, but the driving force and structures that allow for protein functions have not been elucidated at the atomic level for even one protein. We have been able to clarify that the inter-subunit hydrophobic interaction driving the electrostatic opening of the pore in aquaporin 0 (AQP0). Aquaporins are membrane channels for water and small non-ionic solutes found in animals, plants, and microbes. The structures of aquaporins have high homology and consist of homotetramers, each monomer of which has one pore for a water channel. Each pore has two narrow portions: one is the narrowest constriction region consisting of aromatic residues and an arginine (ar/R), and another is two asparagine-proline-alanine (NPA) homolog portions. Here we show that an inter-subunit hydrophobic interaction in AQP0 drives a "stick" portion consisting of four amino acids toward the pore and the tip of the stick portion, consisting of a nitrogen atom, opens the pore: that movement is the swing mechanism (this http URL). The energetics and conformational change of amino acids participating in the swing mechanism confirm this view. The swing mechanism in which inter-subunit hydrophobic interactions in the tetramer drive the on-off switching of the pore explains why aquaporins consist of tetramers. Here, we report that experimental and molecular dynamics findings using various mutants support this view of the swing mechanism. The finding that mutants of amino acids in AQP2 corresponding to the stick of the swing mechanism cause severe recessive nephrogenic diabetes insipidus (NDI) demonstrates the critical role of the swing mechanism for the aquaporin function.

We report first that the inter-subunit hydrophobic interaction in aquaporin 0 drives the electrostatic opening of the aquaporin pore at the atomic level.

**Introduction**

Aquaporins are membrane channels for water and small non-ionic solutes found in animals, plants, and microbes [1]-[3]. The structures of aquaporins are known to have high homology and to consist of homotetramers, each monomer of which has one pore for a water channel [4]-[6]. Each pore has two narrow portions: one is the narrowest constriction region consisting of aromatic residues and an arginine (ar/R), and another is two asparagine-proline-alanine (NPA) homolog portions [4]-[6]. Some fundamental questions remain unsolved: Why do they need to be tetramers, though each monomer owns each pore? Are there inter-subunit interactions? If interactions between subunits are present, are they connected to gating mechanisms? AQP0 has one of the lowest water permeabilities among the aquaporins [2]. In AQP0, Cys189, which is most important for mercury sensitivity, is replaced by Ala [5], [6]. However, AQP4 [7] and plant aquaporins [3], having high water permeability, also do not have the cystine corresponding to AQP1 Cys189. AQP0 has low water permeability but does not have Cys189, suggesting that AQP0 has at least one water pathway essential and common to aquaporins. Furthermore, the structure of AQP0 has been determined with high resolution suitable for atomic-level analysis [6]. For these reasons, we have studied the water permeation mechanisms of AQP0 in detail at the atomic level.

In our observations of molecular dynamics simulation images of a bovine AQP0 (Fig. 1), a nitrogen atom of the asparagine sidechain, $N119_i.n\delta2$ (asparagine 119 i subunit. atom), in the upper stream of the narrowest constriction was bound to an oxygen atom of a glycine, $G180_i.o$, which is one component contributing to the constriction, and, at the same time, the constriction was opened (Fig. 2Avi and 7A)(see a formation of hydrogen bond between $N119_i.n\delta2$ and $G180_i.o$). Furthermore, the $N119_i$ was in one uppermost stream random coil, which contained the histidine, $H122_i$, fourth from the $N119_i$ (Fig. 7A). If the random coil moves as a whole, that is, the stick - $N119_i$, $T120_i$, $L121_i$, and $H122_i$ - move together, the movement of the $H122_i$ can make the far $N119_i$ move toward the $G180_i.o$. In accordance with this reasoning, the $H122_i$ (i chain), must undergo a hydrophobic interaction with a proline, $P109_{i-1}$ (i-1 chain), in the contacting neighboring subunit. Thus, our hypothesis contains a four-fold prediction: (i) the uppermost stream proline, Pro109$_{i-1}$, in the neighboring subunit, i-1, attracts a H122$_i$ in the subunit, i, by hydrophobic interaction, (ii) by their mutual attraction between P109$_{i-1}$ and H122$_i$, the four amino acid strand, the stick, in the same random coil, together moves from the periphery toward the pore region, meaning that (iii) the N119$_i$.nδ2 moves to the pore region, that is, to G180.o, and (iv) the binding of the N119$_i$.nδ2 to the G180.o of the constriction region causes the separation of electrostatic interaction between G180.o, and arginine nitrogen, R187.nη2, of the constriction region, that is, the constriction is open.

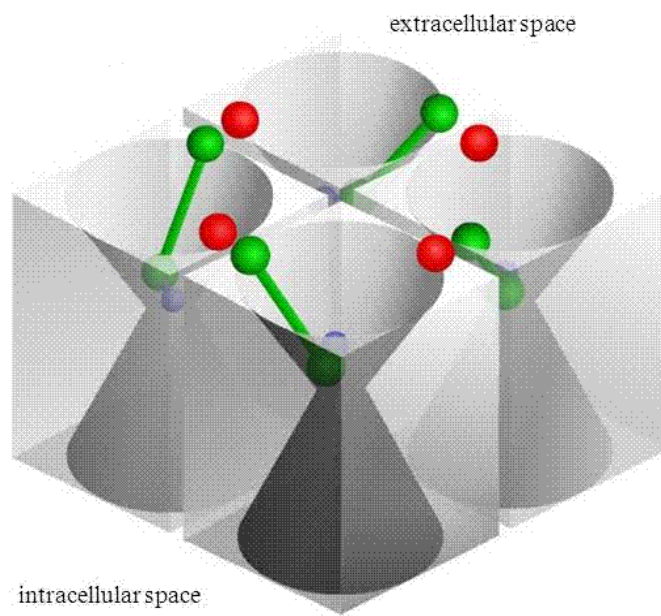

**Figure 1. The simplified illustration of an AQP0.** The aquaporin 0 consists of 4 homotetramers, each monomer of which has one pore for a water channel, like an hourglass. Red ball is $P109_{i-1}$ of another subunit. Green dumb-bell, the stick, consists of $H122_i$ (upper ball), L121i, T120i, and N119i (lower ball), and blue ball, the tip, is $N119i.n\delta2$. See the text about the open-close mechanism of the constriction.

**Results and Discussion**

**Inter-subunit hydrophobic interaction and conformational changes**

In our analyses of 10 ns molecular simulation of wild type bovine AQP0 for clarifying the presence of the "swing mechanisms", at three points (around the first ns in the B chain (Fig. 2Civ), around the fourth ns and 8th to 9th ns in the C chain (Fig. 3Bivc)), the pore first having opened secondly closed and finally opened again (see Fig. 2A and Video S1 for Ancillary video). First, it was confirmed that the inter-subunit hydrophobic interaction between P109$_{i-1}$ and H122$_i$ was present, causing the pore to open. Next, as the inter-subunit hydrophobic interaction became weak, R187.nη2 was bound to G180.o, that is, the pore was closed (Fig.2Ai). The stronger the inter-subunit hydrophobic interaction was, the nearer was the stick N119 to G180 (Fig. 2Aii - Av). This approach of N119 to G180 caused R187.nη2 to separate from G180.o. Further approach of N119 to G180 induced N119.nδ2 to bind to G180.o and this N119.nδ2 binding to G180.o caused R187.nη2 to further separate from G180.o (see the blue line of Fig. 2Civ and Fig. 3Bivc), achieving the pore opening (Fig. 2Avi and 7A and B). Thus, we confirmed the swing mechanism.

Except for the above three points, the pore of the constriction was opened (Fig. 2Civ and Fig. 3Biv). i) The inter-subunit hydrophobic interaction between P109$_a$ and H122$_b$ in the open state (Fig. 2Avi, 7A, and B), measured by accessible surface area (ASA) methods [8], [9], was -1.61 ± 0.44 kcal/mol (Fig. 2Ci). Angle calculation clarified ii) that the movement of H122.cα was conducted to N119.cα (Fig. 2Cii and Ciii). Simultaneously in i) and ii), we observed critical conformational changes in P109$_{i-1}$ and H122$_i$ and especially in H122$_i$, the side chain approached and twisted to face P109$_{i-1}$ (Fig. 4B) with further widening of the angle (Fig. 5A), meaning that considerably strong forces worked between P109 and the H122 side chain (large bent arrow) to make N119.nδ1 move toward the pore (small arrow)(Fig. 4B). These conformational movements of the H122$_i$ side chain against steric hindrance [10] would correspond with most of the energy increase of H122$_i$ (see the following Energetics section). The measurement of distances demonstrated iii) that N119.nδ2 binding to G180.o caused R187.nη2 to separate from G180.o (Fig. 2C); this is the pore opening. These results showed that inter-subunit hydrophobic interactions drove the electrostatic interaction to open the pore (see Fig. 2 and Fig. 4B). These findings concerning the swing mechanisms indicate that a low Arrhenius activation energy underlies the action of the aquaporins [11]-[14].

**Energetics of the swing mechanism**

Energy calculation of each amino acid in wild and mutants and the open-close-reopen phenomena, mentioned above, clarified the following energetics: Energy increases (-1.77 ± 8.41 kcal/mol)(Fig. 4A) of P109$_a$ and H122$_b$ compared with P109$_c$ in H122S (Fig. 3C) and H122$_d$ in P109S, respectively, were suggested to be brought by the formation of the hydrophobic interaction between

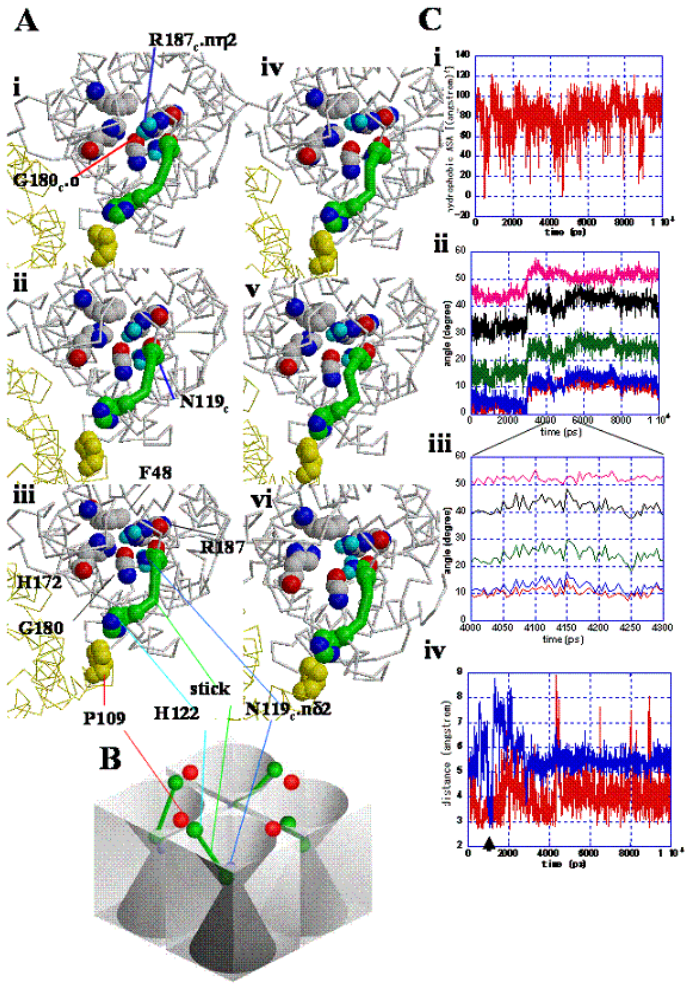
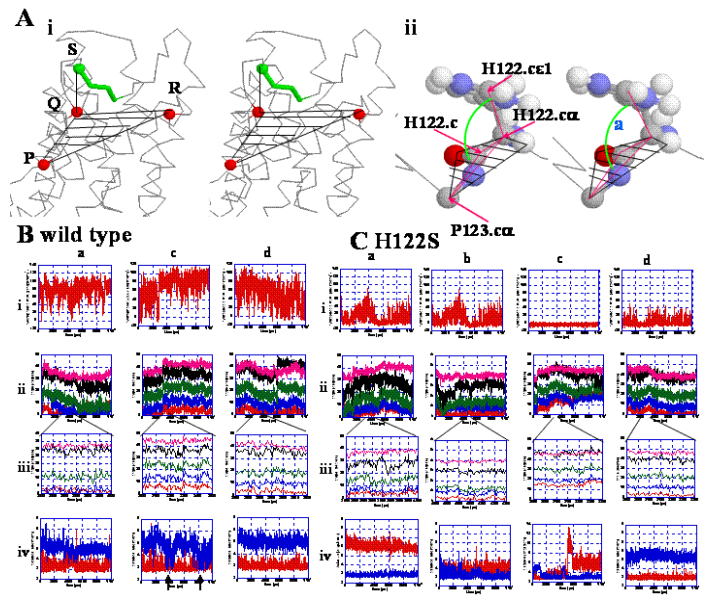

**Figure 2. The swing mechanism opening the pore.**
(**A**) Process of the pore opening in the wild c chain. See the text and the ancillary file (Video S1, Step 1). yellow, b chain (**i**) Snapshot at 4405 ps. (**ii**) 4450 ps structure. (**iii**) 4470 ps structure. (**iv**) 4480 ps structure. (**v**). 4495 ps structure. (**vi**) 4540 ps structure. (**B**) A simplified illustration of an AQP0, the reduced image of Fig. 1. (**C**) The opening mechanism of the constriction in the wild type. (**i**) The inter-subunit hydrophobic interaction between P109$_a$ and H122$_b$ in the open state, measured by accessible surface area (ASA) methods [8,9](Materials and Methods) was 73.02 $\pm$ 20.15 (mean $\pm$ sd) Å$^2$ (calculated by Brunger's X-PLOR [15]), corresponding -1.61 $\pm$ 0.44 kcal/mol (Fig. 2Ci), while, in the original X ray structure, those were not recorded (Fig. 4B, lower portion)(0 Å$^2$ ASA). (**ii**) Movements of the stick were synchronized, except for A25$_b$.c$\alpha$ (Materials and Methods), meaning that they are a connected rigid body [30]. Pink, A25$_b$.c$\alpha$; black, H122$_b$.c$\alpha$; green, L121$_b$.c$\alpha$; blue, T120$_b$.c$\alpha$; red, N119$_b$.c$\alpha$. (**iii**) Synchronized movements in small 5 ps steps of the stick. (**iv**) Distance between N119.n$\delta$2 and G180.o (red line) and between G180.o and R187.n$\eta$2 (blue line). Around a black arrow was the open-close-reopen phenomena observed.

P109$_{i-1}$ and H122$_i$ (Fig. 4B), approximately according to the value calculated by ASA [8], [9]. These energy increases were conducted to L121$_b$ and T120$_b$ and caused N119$_i$.n$\delta$2 to approach to G180.o (Fig. 4C). N119.c$\alpha$ approached to G180.c$\alpha$ with 0.58 Å, as compared to the average structure with X ray structure. In these approaches and the large vibration (10 ~ 20 kcal/mol)(Fig. 4A) of the entire molecule, R187$_b$.n$\eta$2 was separated from G180$_b$.o (Fig. 2A). After that, N119$_b$.n$\delta$2 was attracted toward G180$_b$.o by the electrostatic interaction between N119$_b$.n$\delta$2 and G180$_b$.o and bound to G180$_b$.o (Fig. 2Avi and 4C). That is, it did not need to exceed the energy barrier (Fig. 4C). Thus, actually the inter-subunit hydrophobic interaction was a force that acts an elaborate control to drive the electrostatic interaction to open the pore (Fig. 4B). These phenomena occurred in more than 10 kcal/mol of energy vibration and a few kcal/mol of hydrophobic interaction controlled the swing mechanisms. While the hydrophobic interaction as a force at the macromolecule level has been measured [15], [16], our calculation would make the atomic level analysis of various molecules possible and supply a useful tool for that analysis.

**Figure 3. The Angle calculation and the conformational changes in wild H122 side chain.**
**A.** Angle calculation is indicated by stereo view (see Materials and Methods). (**i**). P represents the L139.c$\alpha$ligand. Q is the V127.c$\alpha$ portion. R is the A101.c$\alpha$ ligand on the aquaporin tetramer molecule. These three points are comparatively immovable $\alpha$ helix. S is the main chain c$\alpha$ to be measured. (**ii**). Angle of H122 side chain was calculated by measurements of four points, P, Q, R, and S. **B.** Swing mechanisms of wild type in a, c, and d chains. (**i**), (**ii**), (**iii**), and (**iv**) are the same as (i), (ii), (iii), and (iv) in b chain of Fig 2C. **C.** Closing mechanisms in a, b, c, and d chains of H122S mutant. (**i**), (**ii**), (**iii**), and (**iv**) are the same as (i), (ii), (iii), and (iv) of B.

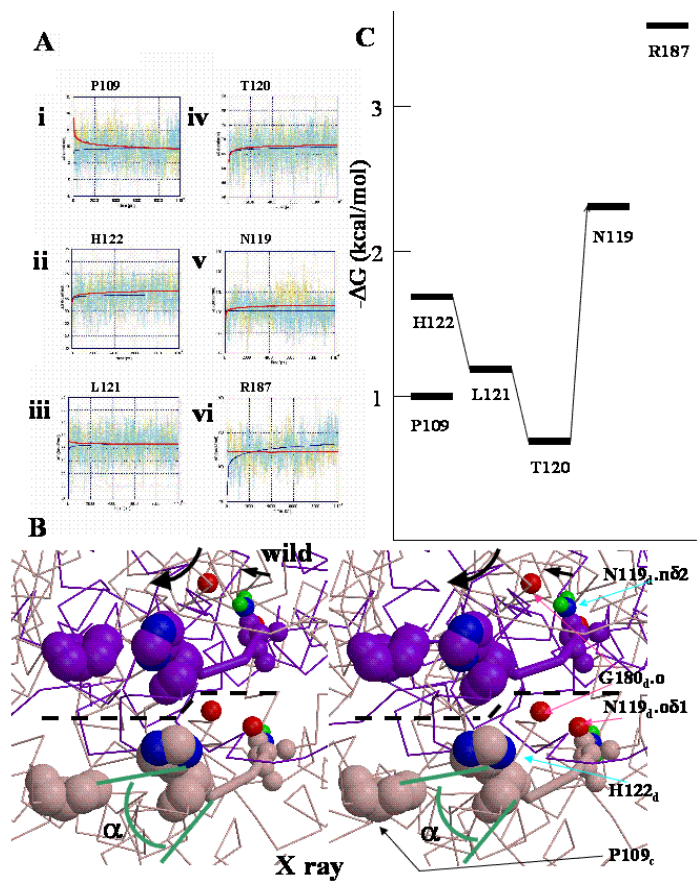

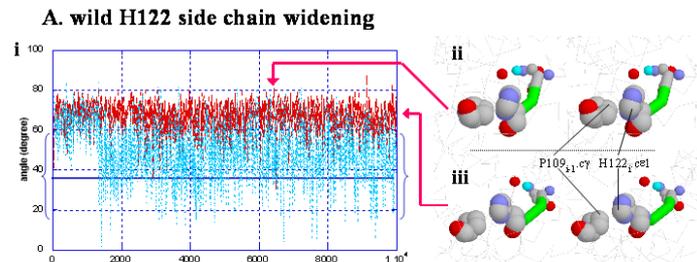

**Figure 4. Energetics of the swing mechanism in the wild type.**
(**A**) Energy increase. (**i**) P109. (**ii**) H122. (**iii**) L121. (**iv**) T120. (**v**) N119. (**vi**) R187. The solid line represents the logarithmic fitting. (**B**) The 4th-10th ns average structure (P109$_c$.c-H122$_d$.cε1, 3.84 Å)(the same as Fig. 2Biii) of the wild type was superimposed and shifted on the original X-ray structure (lower; P109$_{i-1}$.c-H122$_i$.cε1, 6.39 Å). See text. The average 3.84 Å was very close, because the normal limit of C-C nonbonded contact is 3.0 Å and the half thickness in the aromatic group (similar to histidine and proline) is 1.7 Å [30]. Note the N119.nδ2-G180.o hydrogen bonding. (**C**) The potential difference in the swing mechanisms. Molecular images were produced using Rasmol with Molscript transfer.

**Physiological and simulation results of the wild type and mutants**

**Figure 5. The conformational changes in wild H122 side chain. A. (i).** Conformational changes of H122 side chain were observed in the wild d chain (magenta), original X ray (the average, blue bar; the observed range, parenthesis), and P109S (cyan). The inter-subunit hydrophobic interaction of P109i-1 pulling the H122i side chain caused conformational changes of the side chain, the angle difference, 18.40 ± 14.76 (n=1,700; p=0), between wild H122 side chain angle, 66.74 ± 6.89 degree, and P109S H122 side chain angle, 48.07 ± 12.77 (Fig. 4B). Wild H122 side chain angle, magenta; P103S H122 side chain angle, cyan; the original X ray, blue bar and parenthesis. (**ii**). 3.41 Å nearest contact between P109i-1.cγ and H122i.cε1 was found in 6395 ps structure of the wild type (shown by stereo view). (**iii**). 3.84 Å 6th-10th average distance.

For further understanding of the fact that in the swing mechanism the inter-subunit hydrophobic interaction is a cause and the pore opening by electrostatic interaction is a result, we examined mutants prepared both for electrophysiological methods and simulations. Hydrophilic H122S and P109S mutants which were estimated to inhibit water permeability by not forming the inter-subunit hydrophobic interaction demonstrated the decrease of water permeability in electrophysiological methods (Fig. 6B) and showed the constriction closing in simulation methods (Fig. 7C and Fig. 3C), despite the presence of N119 (Fig. 7C). Hydrophobic H122L and P109A mutants which were supposed to maintain water permeability due to forming the inter-subunit hydrophobic interaction kept water permeability in electrophysiological methods (Fig. 6B) and maintained the constriction opening in simulation methods (data not shown). These findings further confirmed that inter-subunit hydrophobic interactions were the cause of the swing mechanism. The N119A mutant not having a nitrogen atom in the A119 side chain (which was estimated to inhibit water permeability) demonstrated a decrease in water permeation in electrophysiological methods (Fig. 6B) and showed the pore closing in the molecular dynamics simulation (Fig. 7D). The N119Q having a nitrogen atom in Q119 which was supposed to maintain water permeability kept water permeability in electrophysiological methods (Fig. 6B) and maintained the pore opening in the simulation (data not shown). Thus, the presence of the nitrogen atom of 119 is indispensable for the pore opening.

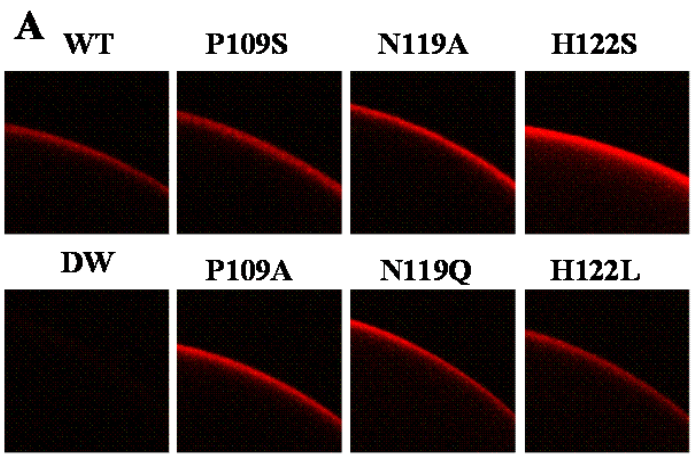

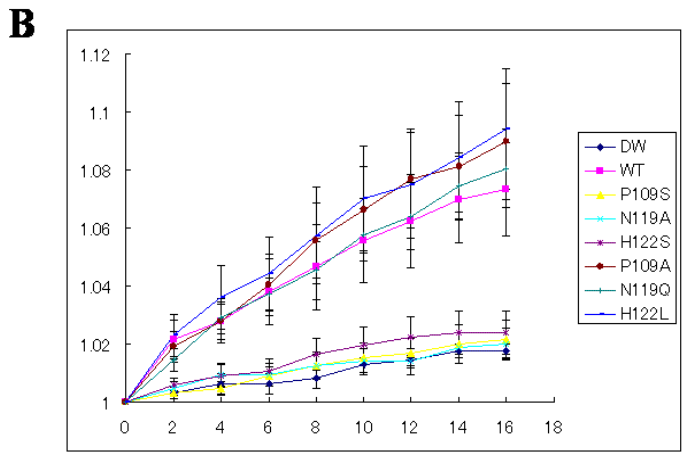

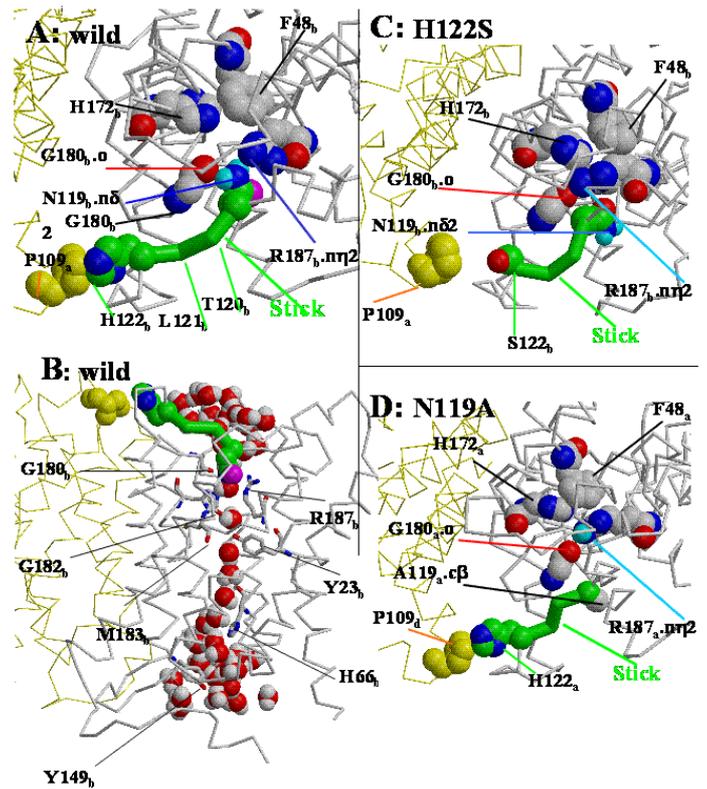

**Figure 6. Cell surface expression of the wild type and mutants and their water permeability measured by physiological methods.**
(**A**) Immunocytochemistry of oocytes injected with cRNA encoding the wild-type AQP0 (WT), P109S, N119A, H122S, P109A, N119Q, H122L, and DW (distilled water). (**B**) Water permeability of the wild type and mutants measured by physiological methods. Water permeation was inhibited in P109S, H122S, and N119A mutants, while water molecules passed through the wild type, P109A, H122L, and N119Q mutants.

**Figure 7. Pore opening (A and B) and closing (C and D).**
(**A**) The 4th-10th ns average structure in the wild type constriction opening by the swing mechanisms, viewed from the extracellular side. (**B**) 8400ps snapshot in the wild type viewed from the lateral side. Note that the R187 side chain nitrogen atoms presumably protonated by separation from G180.o and the G180.o exposed toward the extracellular side (Fig. 7A) attracted water molecules by the electrostatic interaction (Video S1, Step 2). (**C**) The 6th-10th ns average structure of H122S in the constriction closing. Note that N119$_b$.nδ2 faced against the constriction. Pore diameters [33] of H122S, P109S and N119A were narrower by about 0.5 Å than that of wild type, P109A, H122L, and N119Q (compare Fig. 7A, C, and D). (**D**) The 6th-10th ns average structure of N119A in the constriction closing, though the strong Pro109$_d$-H122$_a$ hydrophobic interaction. Note that not only is the pore getting narrow, but also the pore consists of hydrophobic F48, H172 and neutralized G180 and R197 components, which is the same as in the upper C. Water molecules can not pass through the hydrophobic gap, because the movement of an electrostatic molecule through such a gap would require a large amount of energy [34].

The fact that hydrophobic interactions play a critical role in cellular functioning is commonly observed, and suggests that hydrophobic interactions should be included in models of the interaction of macromolecules, e.g., in the "rotation model" [9], [17]-[23].

**Nephrogenic diabetes insipidus.**

T126M in AQP2 which corresponds to the stick L121 in AQP0 causes severe recessive nephrogenic diabetes insipidus (NDI)[24]. T125M corresponding with the stick T120 of AQP0 was also identified in mutations of human AQP2 found in patients with NDI and the water permeability of the oocyte expressing T125M and T126M was lower than that of the oocyte expressing wild-type AQP2 [25], [26]. Furthermore, N123D and N123W of rat AQP2 (corresponds with N119 AQP0) decreased the oocyte water permeability [27]. These findings are explained by the fact that in these mutants the swing mechanism did not work. Since, in not only AQP0 and AQP2 but also many aquaporins, both the upper-most inter-subunit hydrophobic interaction and the asparagine very near the narrowest constriction are present, the swing mechanism is suggested to be common in many aquaporins. However, in order to determine whether the swing mechanism works in a way similar to AQP0 further examinations are needed.

These experiments taken individually and in aggregate established that this bovine AQP homologue contains a gate-opening system which was driven by atom-to-atom interactions, that is, an inter-subunit hydrophobic interaction, electrostatic interactions, etc. Molecular dynamics simulations, which allow for examination of protein performance at the atomic level, have made substantial contributions to understanding the permeation mechanisms for water [14], [28], [29], glycerol, gases, and ions, and the inhibition mechanism of proton transduction [29]. Our findings on the swing mechanism in AQP0 imply that molecular dynamics simulation tools can further elucidate new systems at the atomic level.

**Conclusions**

So far, more than 82,000 protein structures have been reported in the Protein Data Bank, but the driving force and structures that allow for protein functions have not been elucidated at the atomic level for even one protein. We report first that the inter-subunit hydrophobic interaction in aquaporin 0 drives the electrostatic opening of the aquaporin pore at the atomic level.

**Materials and Methods**

**Molecular dynamics simulation (10 ns)**

MD simulation is a widely used and effective method to investigate protein dynamics and their physical properties [22], [23]. X-ray structures of AQP0 from *Bos Taurus* (bAQP0)(Protein Data Bank (PDB) ID: 1YMG)[6], and of P109S, H122S, P109A, H122L, N119A and N119Q mutants prepared on the basis of 1YMG structure were used as the initial structures for our simulations. Each AQP0 and mutants was then placed in a DMPC membrane, which was modified with partial charge calculated using RHF/6-31*G single-point calculation with Gaussian 98 (Gaussian Inc.) and the restrained electrostatic potential method, and was solvated with TIP3P water molecules. After the neutralization, the solvent and counter ions except for water molecules in the pore were optimized by 5,000 steps of energy minimization while the positions of the water molecules in the pore and solute were fixed. Next, the solute and water molecules in the pore were optimized while the position of the counter ions and solvent atoms were kept frozen. All MD simulations were carried out using Amber ver. 7 on personal computers (Pentium III 933 MHz × 32). The bond length involving hydrogen atoms was constrained to equilibrium length by the SHAKE method and the time step was set at 1 fs. Amber parm96, a parameter set for molecular mechanical force fields used for simulations of biomolecules, was adopted. The systems (including ~ 100,000 atoms) were heated to 300K for 0.05 ns and equilibrated for 1.95 ns (total of 2.0 ns) in the NPT ensemble, with periodic boundary conditions and using the particle mesh Ewald method. The temperature and pressure were kept constant at 300K (with a time constant of 1.0 ps) and 1 atm (with a relaxation time of 0.2 ps), respectively.

**Angle calculation.**

Three points in the comparatively immovable α helix, L139.cα (P), V127.cα (Q), and A101.cα (R) were selected and the angle between one line (consisting of one main-chain cα (S) to be measured, and one point, V127.cα (Q)) and the plane constructed using these three points, P, Q and R, where they meet, was measured (Fig. 3Ai). In the case of the histidine side chain, the angle between one line of H122.cε1 (S) and H122.cα (Q) and one plane consisting of three points, H122.cα (Q), H122.c (P) and P123.cα (R) were measured (Fig. 3Aii). In the peptide bond, $c\alpha_i$, $c_i$, $o_i$, $n_{i+1}$, $h_{i+1}$, and $c\alpha_{i+1}$ lie in one plane [30].

**Hydrophobic interactions and energy calculation estimated by ASA [8], [9]**

The ASAs (probe radius 1.4 Å) were calculated using the X-PLOR program [31] based on the concept of atomic accessiblilities [32]. In two contacting molecules, the ASAs (A) of one molecule only, those (B) of another contacting molecule only, and those (C) of one new molecule consisting of the two contacting molecules were estimated by X-PLOR. Buried ASAs (D) of two contacting molecules were calculated as an equation [9] simply, as follows: D = (A + B) - C.

**Energy calculation of each amino acid in molecular dynamics simulation itself.**

As the molecular dynamics simulation is based on energy calculation of atom-atom interactions, we can determine the energy of each amino acid. On the other hand, energy calculations of ASAs can estimate only the contact between noncovalent atoms, but cannot estimate each energy of each amino acid.

**Preparation of average structure.**

PDB file is composed of x, y, and z coordinates of all component atoms. Each x, y, and z coordinate of the atoms in fixed simulation files has the same number as the number of the fixed simulation files. The mean of one x coordinate of one atom in the fixed files was found by adding all of the x coordinate of the atom in the files together and dividing by the number of the fixed files. Each of x, y, and z coordinates was found by the same calculations and an average PDB file was produced.

**Cloning of bovine AQP0 mutants.**

Full-length cDNA of bovine AQP0 (bAQP0) in pXβG-ev1 was kindly presented by Professor Peter Agre, Duke University. Point mutations were introduced into bAQP0 by PCR by using the corresponding mutation primers, and the mutants bAQP0-P109S, bAQP0-P109A, bAQP0-N119A, bAQP0-N119Q, bAQP0-H122S and bAQP0-H122L were generated. The fidelity of all constructs was verified by DNA sequencing.

**Expression in *Xenopus* oocytes and swelling assays.**

After linearization of the bAQP0 constructs with NotI, T3 RNA polymerase was used for reverse transcription (mMESSAGE mMACHINE T3 kit, Ambion, Austin, TX). Oocytes at stages δ-ε were obtained from female *Xenopus laevis*. Each oocyte was injected with either 30 nl of water or 3 ng of wild-type or mutated bAQP0 cRNAs using a Drummond Nanoject microdispensor (Drummond Scientific, Broomhall, PA). The oocytes were incubated for 3 days at 18°C in PS solution containing (mM) 96 NaCl, 2 KCl, 1.8 $CaCl_2$, 1 $MgCl_2$, 5 Hepes, 2.5 pyruvate Na, 0.5 theophylline, supplemented with 100 U/ml penicillin and 100 µg/ml streptomycin, pH 7.6, and then subjected to a swelling assay. Osmotic volume increase of oocytes in response to a 2-fold dilution of PS solution with distilled water was monitored at 20°C under a Nikon TS100 microscope and a CCD camera connected to a computer. Images of the oocyte silhouette were analyzed with ImageJ software, and the oocyte volume was calculated from the cross-sectional area of the oocyte, assuming the oocyte to be a perfect sphere. AQP0 expression at the oocyte surface was checked by immunocytochemistry using a rabbit polyclonal anti-AQP0 antibody (Calbiochem) and a Cy3-conjugated secondary antibody (Jackson ImmunoResearch Lab.).

**Statistics**

We used t-tests with n = 2,000 except where mentioned. Figures were drawn using Kaleida Graph, where a logarithmic retrogression curve was calculated.

**Acknowledgements**

We thank Peter Agre for kindly supplying AQP0, Hiro and Shizuka Goto for the illustrations, and Norman D. Cook for stimulating discussions.

**Author Contributions**

**Competing interests**

The authors declare no competing financial interests.

**Abbreviations**

**AQP0**: aquaporin 0; **AQP1**: aquaporin 1; **NDI**: nephrogenic diabetes insipidus; **A**: alanine; **E**: glutamate; **Q**: glutamine; **D**: aspartate; **N**: asparagine; **L**: leucine; **G**: glycine; **K**: lysine; **S**: serine; **V**: valine; **R**: arginine; **T**: threonine; **P**: proline; **I**: isoleucine; **M**: methionine; **F**: phenylalanine; **Y**: tyrosine; **C**: cysteine; **W**: tryptophan; **H**: histidine

**Ancillary file**

We put an ancillary file in the site: http://www.apph.tohoku-gakuin.ac.jp/iwamoto/video_S1.gif , as follows;

A still image from video_S1 file

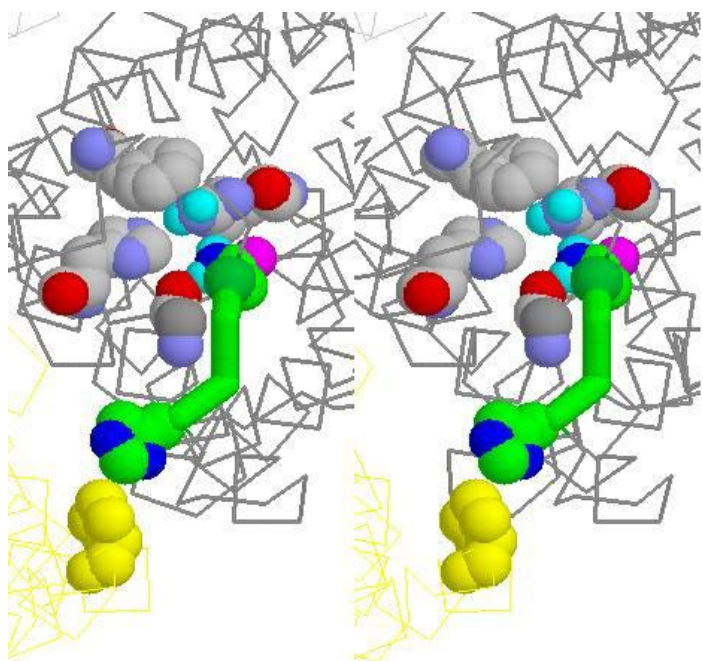

Title of data: Video_S1 legend. The opening, closing and re-opening of the pore and the water permeation. Description of data: Display Method: Immediately after clicking the left mouse button, you can see the video. Contents: 1. Step 1. Around the fourth ns in the wild c chain. yellow, b chain. First, the pore opens, as the inter-subunit hydrophobic interaction between P109 and H122 is formed. Second, as the inter-subunit hydrophobic interaction becomes weak, the pore gets closed. Gradually, the hydrophobic interaction increases, and the pore gets the wider. Finally, when the hydrophobic interaction is formed, the pore opens. See Fig. 2A for a description of the images. Step 2. Around 8500 ps in the wild d chain, there occurs the permeation of one water molecule (small white-red-white ball) through the pore. The water molecule is first seen on the near side, passes through the constriction, and moves to the far side.